\documentclass[aps,pra,reprint,superscriptaddress]{revtex4-1}

\usepackage{soul}
\usepackage{graphicx}
\usepackage{amsmath,amssymb,amsfonts}
\usepackage{enumitem}
\usepackage{xcolor}
\usepackage{bm}
\usepackage{soul,xcolor}
\usepackage{tikz,tkz-euclide}
\usepackage[colorlinks=true,citecolor=blue,linkcolor=blue,urlcolor=blue]{hyperref}
\usepackage[section]{placeins}

\begin{document}

\begin{abstract}
The nodal surfaces of the many-body wavefunction are fundamental geometric features that encode critical information regarding particle statistics and  {their} interaction. Directly probing these structures, particularly in  correlated quantum systems, remains a significant experimental challenge. Here, we provide rigorous results on the structure of the many-body wavefunction and propose to use an interferometric technique to probe its zeros in ultra-cold atomic systems. Specifically, we refer to the so-called heterodyne interferometric reconstruction of the phase of the many-body wavefunction. We prove that the sought nodal surfaces show up as specific discontinuities in the interference fringes. Following Leggett, both `symmetry-dictated' nodal surfaces, due to particle statistics, and `non-symmetry dictated' nodal surfaces emerging from interaction effects, can be probed. We demonstrate how the spin degrees of freedom, effectively modifying the structure of the nodal surfaces of the many-body wavefunction, leave distinct fingerprints in the resulting interference pattern. Our work addresses important features of the structure of the many-body wavefunction that are broadly relevant for quantum science ranging from conceptual aspects to computational questions of extended systems and quantum simulation.
\end{abstract}

\title{
Interferometric probe for the zeros of the many-body wavefunction
}

\author{Wayne J. Chetcuti}
\affiliation{Université Grenoble-Alpes, CNRS, LPMMC, 38000 Grenoble, France}
\affiliation{Quantum Research Centre, Technology Innovation Institute, Abu Dhabi, UAE}
\author{Anna Minguzzi}
\affiliation{Université Grenoble-Alpes, CNRS, LPMMC, 38000 Grenoble, France}
\author{Juan Polo}
\affiliation{Quantum Research Centre, Technology Innovation Institute, Abu Dhabi, UAE}
\author{Luigi Amico}
\affiliation{Quantum Research Centre, Technology Innovation Institute, Abu Dhabi, UAE}
\affiliation{Dipartimento di Fisica e Astronomia, Via S. Sofia 64, 95127 Catania, Italy}
\affiliation{INFN-Sezione di Catania, Via S. Sofia 64, 95127 Catania, Italy}
\maketitle
\date{\today}

{\textbf{Introduction --}}  Quantum many-body wavefunctions describe collective quantum states of  systems composed of a large number of  particles. They are  high-dimensional functions of the $N_p$ particle coordinates and intrinsic degrees of freedom (e.g., spin). Particle-particle interactions generically give rise to non-trivial patterns of quantum correlations~\cite{amico2008entanglement,eisert2010colloquium}
that are both bedrock for  emergent phenomena in condensed matter physics, ranging from superconductivity and superfluidity to topological or other quantum phases of matter~\cite{keimer2017physics}, and provide  key  resources  to be harnessed by  quantum technology~\cite{acin2018quantum}. As a fundamental constraint on their form, many-body wavefunctions need to be anti-symmetric or symmetric, for fermionic or bosonic particles, respectively. A full reconstruction of a generic quantum many-body state remains a formidable task \cite{ceperley1991fermion, gross2010quantum,torlai2018neural,orus2019tensor,kokail2021entanglement,lanyon2017efficient}. Profound insights into the fundamental structure of many-body wavefunctions can be gained  through an argument originally due to Leggett~\cite{leggett1991theorem}. The argument posits that for a system of interacting (spinless) fermions  on a ring pierced by a magnetic field, the Pauli principle implies that the many-body wavefunction displays specific nodal surfaces (in the $N_{p}$-dimensional coordinate space) in which it vanishes. Besides such `symmetry-dictated nodal surfaces' arising from particle statistics, other `non-symmetry-dictated nodal surfaces'  may occur, as a result of  combined effects of orbital nodes, interactions, and external fields. For bosons, nodal surfaces are predicted to `recombine' {with} one another to give rise to a smooth superfluid order~\cite{leggett1970can}. 
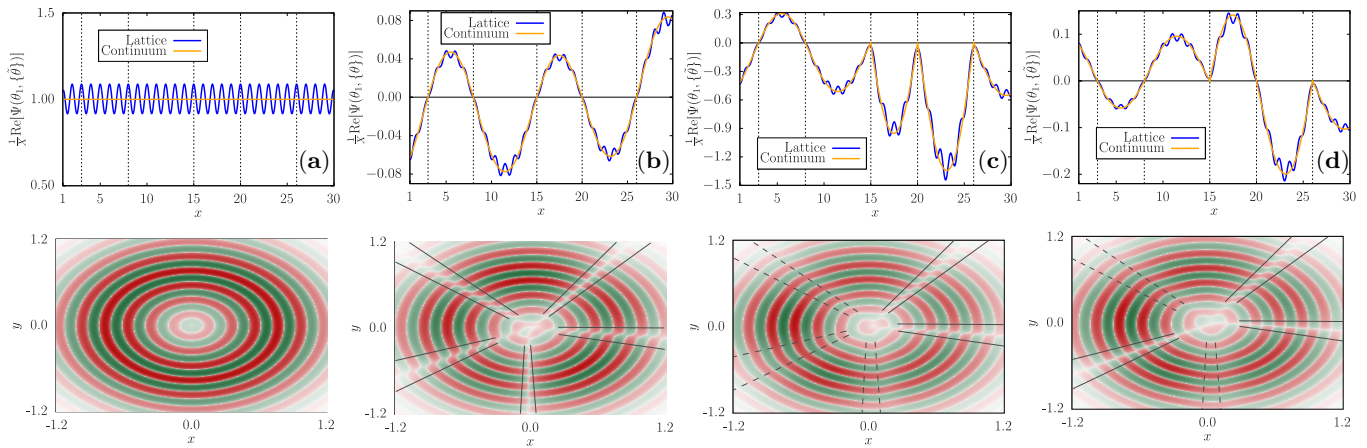
\begin{figure*}[htbp!]
    \centering
    \includegraphics[width=\linewidth]{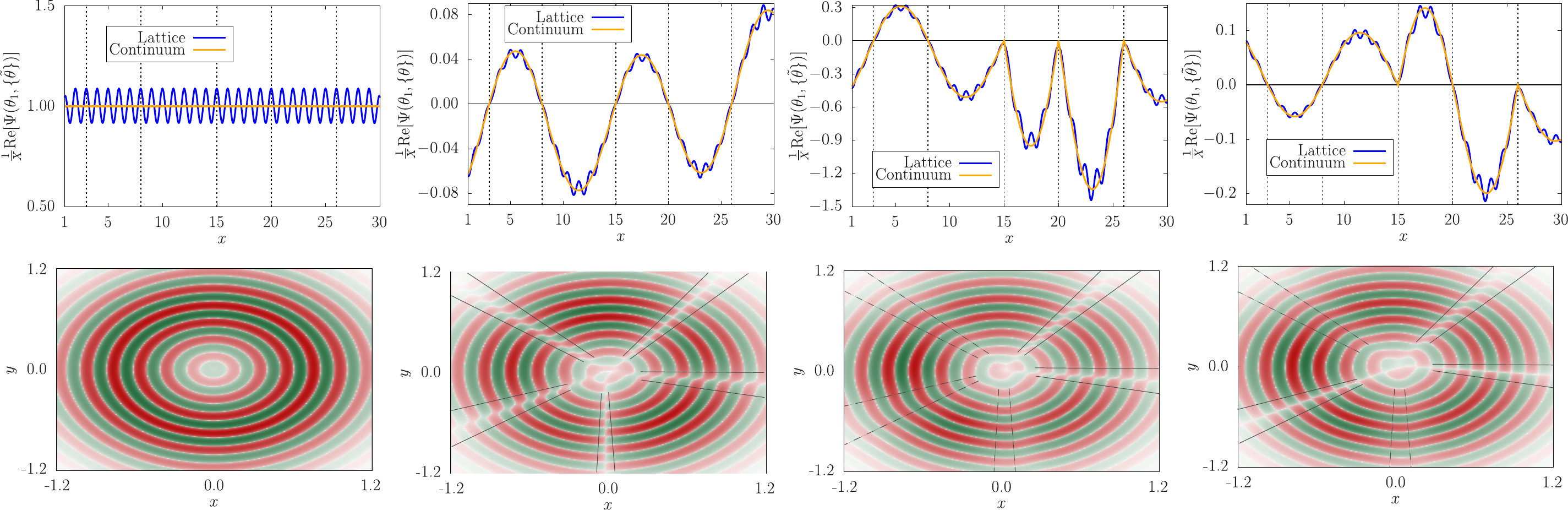}%
    \put(-400,105){(\textbf{a})}
    \put(-272,105){(\textbf{b})}
    \put(-143,105){(\textbf{c})}
    \put(-15,105){(\textbf{d})}
    \caption{\textit{Single-shot protocol.} Top row depicts the one-dimensional wavefunction $\frac{1}{X}\Psi(\theta_{1},\{\tilde{\theta}\})$ in the continuum (orange) and lattice (blue)~\cite{WannierGaussianNote} acquired by fixing $N_{p}-1$ particle coordinates for $N_{p}=6$ \textbf{(a)} non-interacting bosons, \textbf{(b)} spinless fermions and \textbf{(c)},\textbf{(d)} two-component fermions in the limit of strong interactions (exact Bethe Ansatz calculation) [see Sec.~\ref{sec:expect-single-shot} in Supplemental material for the details on the factor $X$]. The bottom row shows the corresponding single-shot interference images Re[$\psi_{1,center}^{*}(\mathbf{x};t)\Psi_{1,ring}(\mathbf{x};t)$. Dislocations (highlighted by a pair solid lines) in the interferograms correspond to nodes in the wavefunction whilst cusps (marked by two dashed lines) appear as deformations of the interference fringes. For a winding number $\ell =0$, there are \textbf{(a)} zero, \textbf{(b)} five and \textbf{(c)} two (three) dislocations (cusps), whilst for $\ell = 1/2$ there are three (two) dislocations (cusps) present. For all instances, the particle coordinates are fixed as $\{\theta_{2},\theta_{3},\theta_{4},\theta_{5},\theta_{6}\} = 3, 8, 15, 20, 26$ indicated by the dotted lines in the first row.  In the plots, the Wannier functions $w_{\theta_l}(\mathbf{x}_j)$ are taken as two-dimensional time-evolved Gaussians functions; the parameters are the radius $R=1.5$, number of sites $N_{s}=30$, the Gaussian width $\sigma = 0.15$ and time of the expansion $\omega_{0}t=1.5$. The color bar is taken to be non-linear by setting $\mathrm{sign}(Y)|Y)|^{\frac{1}{4}}$, with $Y$ denoting the quantity being plotted. 
    } 
    \label{fig:singleshotnoninteracting}
\end{figure*}

Here, we prove  that the nodal surfaces of the many-body wavefunction leave specific signatures in suitable interference protocols. We refer to a cold atom implementation in which the matter-wave is trapped in ring circuits~\cite{amico2022atomtronic}. In this context, a specific {self-heterodyne} experimental protocol has been carried out in which a rotating matter-wave in a ring track co-expands with a non-rotating condensate placed in the center of the ring.  The interference can be obtained through the measured density  
$n(\mathbf{x};t)=n_{disk}(\mathbf{x};t)+ n_{ring}(\mathbf{x};t) + 2 \sqrt{n_{disk}(\mathbf{x};t) n_{ring}(\mathbf{x}; t)}\cos{[\delta \xi (\mathbf{x};t)]}$, in which  $\delta\xi$ is the phase difference between the ring and disk condensates~\cite{eckel2014interferometric}. 
Remarkably, {through} such a protocol the phase of a rotating condensate, and in turn, the angular momentum of the particles flowing in the ring, can be detected, the disk {acting as} its reference~\cite{mathew2015self}. The scheme has been successfully implemented also for fermionic systems in the BCS-BEC crossover~\cite{delpace2022imprinting}.

In this work, we demonstrate how the analysis of such interference patterns can probe the nodal surfaces of the many-body  wavefunction. To this end, our analysis focuses on comparing the response of bosonic and fermionic many-body ground-states on a ring in presence of an effective magnetic field. Because of the latter, a winding number is imparted to the wavefunction that, in turn, sustains a persistent current~\cite{polo2025persistent}. Then, the key feature to be relied on is a certain set of line discontinuities, hereafter referred to as \textit{dislocations}, that appear in the aforementioned interferograms. We shall see how such dislocations are indeed phase slips of the many-body wavefunction and mark its nodal surfaces. Non-symmetry dictated nodal surfaces show up as dislocations in the interferogram as result of the combined effect of interaction and winding number. 

To be concrete, we refer to a system of particles localised in a one-dimensional ring. The single-shot interference is constructed for both non-interacting theories and systems with a local infinite particle-particle repulsion. We use these cases to directly prove {analytically} that dislocations arise from a discontinuity in the argument of the many-body wavefunction corresponding to symmetry dictated nodal surfaces. Importantly, the same discontinuity is shown to appear in the atom shot-noise, as quantified by density–density correlators in the absorption images of the  system. Subsequently, we consider full fledged  correlated many-body systems. For this purpose, a two-component (effective spin-$\frac{1}{2}$) Fermi–Hubbard model on a ring pierced by a magnetic field is examined through DMRG simulations~\cite{itensor} and Bethe Ansatz exact methods. We prove how the spin degrees of freedom change the structure of the many-body wavefunction's nodal surfaces, thereby introducing characteristic features into the interference pattern. 

{\textbf{Non-interacting systems --}} In our formalism, the disk is  concentrated in the ring center  $\mathbf{x}_c$. Quantum particles residing in a ring (of $N_{s}$ sites) and in the center are initially trapped and described by $\Psi (\{\mathbf{x}\},\mathbf{x}_c;0|\phi)=\Psi_{ring} (\{\mathbf{x}\};0|\phi)\psi_{center}(\mathbf{x}_c;0)$, where we used the shorthand notation {$\{\mathbf{x}\} \equiv \{\mathbf{x}_1,\dots \mathbf{x}_{N_p}\}$}. The many-body wavefunction {in the presence of a magnetic flux reads as}
$\Psi_{ring} (\{\mathbf{x}\};0{|\phi})=
 {e^{2\pi\imath\frac{\phi}{\phi_{0 }N_{s}}\sum_{l}\theta_{l}}} {\mathcal{A}} [ \psi(\mathbf{x}_j;0)]$
where ${\mathcal{A}}[ \psi(\mathbf{x}_j;0)] $ denotes Slater determinant or  permanent of single-particle wavefunctions $\psi(\mathbf{x}_j;0)$, respectively in the fermionic or bosonic case, each located 
on the {ring's} lattice sites with {positions} $\{\theta_1,...,\theta_{N_p}\}\equiv\{\theta\}$. The effective magnetic field denoted by $\phi$ in units of the flux quantum $\phi_0$, imparts a quantized angular momentum per particle (winding number) $\ell$ to minimise the system's energy. After switching off the traps, the quantum gases in the ring and center undergo a free expansion and interfere, in the aforementioned  self-heterodyne interference protocol~\cite{eckel2014interferometric}. The expansion of many-body ring wavefunction reads 
\begin{equation} 
\Psi_{ring} (\{\mathbf{x}\};t|\phi) = \sum\limits_{\{\theta\}}\Psi_{ring} (\{\theta\};0|\phi)\prod\limits_{j = 1}^{N_{p}}w_{\theta_j}({\mathbf{x}}_j;t) \;,
\label{eq:expandedring}
\end{equation}
in which $\Psi_{ring} (\{\theta\};0|\phi)= {e^{2\pi\imath\frac{\phi}{\phi_{0}N_{s}}\sum_{l}\theta_{l}}} {\cal A} [ \psi(\theta_l;0)]$ and the time evolved single-particle Wannier functions basis $w_{\theta_j}(\mathbf{x}_j;t)\doteq w(\mathbf{x}_{j}-R_{\theta_j};t)$ has been employed (see Sec.~\ref{sec:timeevol} in Supplementary material for explicit calculations). 

In a {\it single-shot experiment}, the interference arises from $\Psi_{1,ring} (\mathbf{x};t) \psi^*_{1,center}(\mathbf{x},t
)$ where the one-particle ring wavefunction $\Psi_{1,ring}  (\mathbf{x};t)$ is identified  by integrating out all the  particles'  coordinates ${\mathbf{x_j}}$ but one in  the density associated to Eq.~\eqref{eq:expandedring}:
\begin{equation}\label{eq:singleparticles}
{\Psi}_{1,ring}(\mathbf{x};t|\phi)=\sum_{\theta_{1}}^{N_{s}}w_{\theta_1}(\mathbf{x};t)\Psi(\mathbf{\theta}_1,\{\tilde{\theta}\}|\phi), 
\end{equation}
where $\{\tilde{\theta}\}=\theta_2,\dots\theta_{N_p}$ are taken to be fixed coordinates in the ring (see Sec.~\ref{sec:expect-single-shot} of the Supplementary material for details).

The interferograms are curves of constant phase $\xi(\mathbf{x};t|\phi)\!=\!\arg\left [\Psi_{ring} (\mathbf{x};t|\phi) \psi^*_{center}({\mathbf{x}};t
|\phi)\right ]$ 
in polar coordinates~\cite{mathew2015self}. Taking the argument of $\psi^*_{center}(\mathbf{x};t
)$ as reference for the phase, $\xi(\mathbf{x};t|\phi)$ reads: 
\begin{equation}
\xi(\mathbf{x};t|\phi
)=\tan^{-1}\left [{{\sum_{\theta_1} Z({\theta}_1,\mathbf{x};t|\phi) \sin \Xi({\theta}_1,\mathbf{x};t|\phi)}\over{\sum_{\theta_1} Z({\theta}_1,\mathbf{x};t|\phi) \cos \Xi({\theta}_1,\mathbf{x};t|\phi)}} \right ],
\label{eq:spiralwave}
\end{equation}
where  $Z ({\theta}_1,\mathbf{x};t|\phi)\!=\!|w_{\theta_1}(\mathbf{x};t) \Psi({\theta}_1,\{\tilde{\theta}\} ;0|\phi)|$ and $\Xi({\theta}_1,\mathbf{x};t|\phi)\!=\!\arg[w_{\theta_1}(\mathbf{x};t)]+ \arg[ \Psi({\theta}_1,\{\tilde{\theta}\};0|\phi)]$. The continuity properties of  $\xi(\mathbf{x};t|\phi)$ are encoded in $\Xi({\theta}_1,\mathbf{x};t|\phi)$. While $\arg[w_{\theta_1}(\mathbf{x};t)]$ are smooth functions of the particles' position on the ring, the analytic properties of $\arg\left [\Psi ({\theta_1},\{\tilde{\theta}\};0|\phi)\right]$ in the ground-state depend on whether the particles are bosons of fermions. Concerning bosonic wavefunctions in homogenous rings,  $\arg\left [\Psi ({\theta_1},\{\tilde{\theta}\};0|\phi) \right ]$ is a smooth function of the particle coordinates on the ring -- Fig.~\ref{fig:singleshotnoninteracting}\textbf{(a)}. For fermionic systems, instead, $\arg\left [\Psi({\theta_1},\{\tilde{\theta}\};0|\phi) \right ]$ displays  discontinuities specifically at the  nodal surfaces of the wavefunction where pairs of particles $(\theta_1,\theta_j)$ can swap their mutual position (see Sec.~\ref{sec:wavefunc} in Supplemental material). {\it{It is because of such discontinuities that dislocations show up in the {interferograms}}} -- Fig.~\ref{fig:singleshotnoninteracting}\textbf{(b)}. 

In the many-body formalism, the interference is reproduced as a suitable expectation value of the relevant physical observables. Clearly, such approach relies on the assumption that physical results can be obtained by averaging on ensembles of the physical system that, experimentally, can be obtained by repeating the expansion experiment multiple times. For the co-expansion experiment under discussion, rather than expectation values of the density, the spiral interference can be obtained through the density-density correlators~\cite{haug2018readout,pecci2021probing,chetcuti2022interference} $G(\mathbf{x},\mathbf{x'};t) = \langle n(\mathbf{x};t)n(\mathbf{x'};t)\rangle$, with $n(\mathbf{x};t) = \Psi^{\dagger}(\mathbf{x};t)\Psi(\mathbf{x};t)$ being the density operator, and where $\Psi^{\dagger}(\mathbf{x};t) = {\Psi}^{\dagger}_{ring}(\mathbf{x};t)+{\psi}^{\dagger}_{center}(\mathbf{x};t)$ are field creation operators for the ring-center system. $G(\mathbf{x},\mathbf{x'};t)$, with one of the coordinates, $\mathbf{x}$ or $\mathbf{x'}$, kept fixed and at intermediate times $t$ has been demonstrated to generate interferograms for a variety of different systems~\cite{haug2018readout,pecci2021probing,pecci2022single,chetcuti2022interference,osterloh2023exact}. In the following, we prove that dislocations in  $G(\mathbf{x},\mathbf{x'};t)$ are indeed caused by the very same physical mechanism discussed for the single-shot measurement based on the wavefunction.

As mentioned previously, the spiral interferogram comes from the ring-center interference terms. Under the assumption that the ring and center are separated at time $t=0$, the cross-correlator factorizes as  
$G_{ring,center}(\mathbf{x},\mathbf{x'};t) \!=\! \rho_{1}^{(ring)}(\mathbf{x},\mathbf{x'};t)\rho_{1}^{(center)}(\mathbf{x'},\mathbf{x};t)$ with $\rho_{1}^{(ring/center)}(\mathbf{x},\mathbf{x'};t)$ being the one-body correlator of the ring/center. In first quantization, the ring one-body density correlator is $\rho_{1}^{(ring)}(\mathbf{x},\mathbf{x'};t) \!=\! \int d\{\mathbf{\tilde{x}}\}\Psi_{ring}^{*}(\mathbf{x},\{\mathbf{\tilde{x}}\};t|\phi)\Psi_{ring}(\mathbf{x'},\{\mathbf{\tilde{x}}\};t|\phi)$,
where we used the notation $\{\mathbf{\tilde{x}}\}\equiv\mathbf{x}_{2},...,\mathbf{x}_{N_{p}}$.
Plugging the expression for the wavefunction in Eq.~\eqref{eq:expandedring}, we can express the ring correlator as
\begin{eqnarray}\label{eq:one-particle-density}
    \rho_{1}^{(ring)}(\mathbf{x},\mathbf{x'};t) = \sum\limits_{\{\tilde{\theta}\}=1}^{N_{s}} {\Psi}^{*}_{1,ring}(\mathbf{x'};t|\phi){\Psi}_{1,ring}(\mathbf{x};t|\phi).
\end{eqnarray}
Since ${\Psi}_{1,ring}(\mathbf{x};t|\phi)$ is given by the same Eq.~\eqref{eq:singleparticles}, this demonstrates that the same argument of the wavefunction  ${\Psi}_{ring}(\mathbf{x};t|\phi)$ entering in $\xi(\mathbf{x};t|\phi)$ is the one that is responsible for establishing the interference in $\rho_{1}^{(ring)}(\mathbf{x},\mathbf{x'};t)$. {\it Consequently, dislocations in $\rho_{1}^{(ring)}(\mathbf{x},\mathbf{x'};t)$, and then in the whole $G_{ring,center}(\mathbf{x},\mathbf{x'};t)$, originate from the very same discontinuities and nodal surfaces we discussed above for the many-body wavefunction.} 

{\textbf{Interacting many-body systems --}} Next, we consider
two-component repulsively interacting fermionic systems. In this case, nodes of the many-body wavefunction still arise due to symmetry upon exchange of particles belonging to the same component. However,  additional nodes or cusps may appear upon exchanging particles belonging to different components due to strong repulsive interactions. Specifically, a system of $N_{p}$ fermionic particles with two-components/colours residing in a one-dimensional ring-shaped optical lattice composed of $N_{s}$ sites is considered, as described by the Fermi-Hubbard model~\cite{lieb1968absence,essler2005one} 
\begin{align}\label{eq:Hamiltonian_lattice}
    \mathcal{H} = \sum\limits_{j=1}^{N_{s}}\bigg[-J\!\!\!\sum\limits_{\alpha=\{\uparrow,\downarrow\}}\!( e^{\imath\Omega}c_{j,\alpha}^{\dagger}c_{j+1,\alpha} + \mathrm{h.c.}) + Un_{j,\uparrow}n_{j,\downarrow} \bigg], 
\end{align}
where $c_{j,\alpha}^{\dagger}$ ($c_{j,\alpha}$) creates (destroys) a fermion with colour $\alpha$ on site $j$ and $n_{j,\alpha}=c_{j,\alpha}^{\dagger}c_{j,\alpha}$ is the local number operator. The parameter $J$ denotes the hopping amplitude, whilst $U\!>\!0$ corresponds to repulsive interaction strengths. 
In the limit of  dilute filling fractions and vanishing lattice spacing, Eq.~\eqref{eq:Hamiltonian_lattice} tends to Gaudin-Yang  models governing fermions residing in the continuous space and  subjected to delta interaction~\cite{essler2005one,guan2013fermi}.
Besides, the thorough applications of the Hubbard model in strongly correlated electronic systems (see f.i.~\cite{baeriswyl1995hubbard}), we comment that two-component Fermi systems with tunable interaction are realized through a variety of cold atoms as $^6$Li or $^{40}$K~\cite{esslinger2010fermi,mazurenko2017cold,tarruell2018quantum,brown2019bad}. Both the lattice and continuous two-component models are integrable by Bethe Ansatz~\cite{lieb1968absence,gaudin1967un,yang1967some}. This feature allows us to access the exact eigenstates of model~\eqref{eq:Hamiltonian_lattice}, labeled by suitable quantum numbers~\cite{essler2005one}.

The complex hopping amplitudes, $J\rightarrow Je^{\imath\Omega}$ known as Peierls substitution~\cite{peierls1933zur}, takes into account the effective magnetic flux $\Omega \! =\! 2\pi\phi/(N_{s}\phi_{0})$. This way, the system sustains quantized matter-wave persistent currents, defined as $I(\Omega) \! =\! -{\partial E_{0}(\Omega)}/{\partial\Omega}$ with $E_{0}$ denoting the ground-state energy. Fermionic systems are characterized by a parity effect: systems with $N_p =4\nu +2$ are diamagnetic whilst systems with $N_p =4\nu$ are paramagnetic, $\nu$ being a positive integer~\cite{waintal2008persistent,chetcuti2022persistent}. Persistent currents of Eq.~\eqref{eq:Hamiltonian_lattice} display a non-trivial dependence on the particle interaction reflected in energy crossings occurring at specific values of $\phi/\phi_0$. In each segment of the resulting piece-wise persistent current  landscape, the many-body wavefunction is characterized by a distinct fractional winding number $\ell$ acquired by the system to counterbalance the applied flux. For model~\eqref{eq:Hamiltonian_lattice}, $\ell$ can be extracted through the Bethe Ansatz charge quasimomenta: $\sum_{n}k_n=\ell$ ($k_n$ are fixed in terms of the quantum numbers labeling the Hamiltonian eigenstates)~\cite{chetcuti2022persistent,polo2025persistent}. At strong repulsion, the periodicity of the persistent current is characterized by a fractionalized flux quantum $\frac{\phi_{0}}{N_{p}}$~\cite{chetcuti2022persistent,osterloh2023exact,polo2024static}. Note that the `total spin' is a constant of the motion of the Hubbard model:  $[\mathcal{H},S^2]=0$, the eigenvalues of $S^2$ are $S(S+1)$, with ${S^{a}}=(1/2)\sum_{j}^{N_{s}}\sum_{\alpha,\beta}c^\dagger_{j,\alpha}(\sigma^{a})_{\alpha}^{\beta}c_{j,\beta}$, $\sigma^{a}$,  $a=\{x,y,z\}$ being Pauli matrices. The analysis shows that depending on the diamagnetic (paramagnetic) parity, the system is in a spin singlet $S=0$ for $\ell = m$ ($\ell = \frac{m+1}{2}$) with integer $m$, whilst for any other $\ell$ the many-body wavefunction transforms as a spin triplet $S=1$~\cite{osterloh2023exact,polo2024static}. 

Firstly, we consider generic situations corresponding to intermediate interactions; then shift focus on the case of very strong interactions amenable to exact analysis~\cite{ogata1990bethe,essler2005one}. Our DMRG numerics for the interferograms are presented in Fig.~\ref{fig:DMRG} (see Sec.~\ref{sec:numerics} in Supplemental material). We find that the  number of dislocations depends on the combination of winding number and interaction. There are  $\frac{N_{p}}{2}-1$ ($\frac{N_{p}}{2}$) dislocations for $\ell \!= \! 0$/$\ell\!=\!1$ ($\ell \!=\! \frac{1}{2}$) for diamagnetic (paramagnetic) parity. 

In the limit of strong interactions as $U\rightarrow\infty$, the Bethe Ansatz equations for charge and spin degrees of freedom decouple~\cite{ogata1990bethe}. Essentially, the model separates into a spinless fermionic Hamiltonian and the $\rm{XXX}$ Heisenberg model accounting for the charge and spin respectively. Consequently, the many-body wavefunction of the system reads~\cite{essler2005one}
\begin{equation}\label{eq:wavestrong}
    \Psi_{ring}^{\infty}(\{\bar{\theta}\},\{\bar{\alpha}\}|\bar{k},\bar{\lambda}) = \langle\bar{\alpha} Q|\bar{\lambda}\rangle \mathrm{det}[e^{\imath k_{m}\theta_{n}}]_{m,n=1}^{N_{p}},
\end{equation}
in which the charge quasimomenta $\bar{k}$ entering the determinant are dependent on the so-called charge and spin quantum numbers denoted by $I_{n}$ and $J_{\alpha}$ respectively via the relation $k_{n} = \frac{2\pi}{N_{s}}\big[I_{n}+\frac{1}{N_{p}}\sum_{\alpha}^{M}J_{\alpha}+\frac{\phi}{\phi_{0}}\big]$~\cite{osterloh2023exact}. In the presence of a flux, the quantum number configurations $\{I_{n},J_{\alpha}\}$ can change to minimize the system's energy at a given winding number $\ell$~\cite{chetcuti2022persistent,chetcuti2023probe}. The spin amplitudes $\langle\bar{\alpha} Q|\bar{\lambda}\rangle$ in Eq.~\eqref{eq:wavestrong} correspond to suitable eigenstates of the Heisenberg  model $H_{\rm{XXX}}$ labeled by spin rapidities $\bar{\lambda}$ in a given coordinate sector $\bar{\alpha} Q$, fixed by the spin quantum numbers $J_{\alpha}$ at a given $\ell$. Through the spin amplitudes, the \textit{nodes of the orbital many-body wavefunction are converted into cusps}, which manifest not as dislocations but as smooth deformations in the single-shot interferogram -- Fig.~\ref{fig:singleshotnoninteracting}\textbf{(c)},\textbf{(d)}. As a result, the number of dislocations are very dependent on the system's winding number -- as it corresponds to a specific Heisenberg eigenstate characterized by a specific set of spin quantum numbers. Compared with non-interacting systems, for  $U\rightarrow\infty$ the number of dislocations in the interference pattern is equal or larger, with the actual number depending on the winding number that can make  nodes into cusps in the wavefunction (see Fig.~\ref{fig:singleshotnoninteracting}).

Summarizing: for $\ell = 0$ ($\ell=\frac{1}{2}$) both numerics and analytics at intermediate and strong interactions confirm the the number of dislocations correspond to symmetry-dictated zeros present in the many-body wavefunction arising from fermionic statistics, for diamagnetic (paramagnetic) systems. Conversely, going to $\ell=\frac{1}{2}$ ($\ell = 0$) the number of dislocations differs marking the presence of a non-symmetry dictated nodal surface emerging due to interactions. It is important to note that the visibility of the dislocations in the single-shot mechanism is heavily reliant on $\langle\bar{\alpha} Q|\bar{\lambda}\rangle$ and in turn on the spin configuration in a given coordinate sector $Q$. 
\begin{figure}[h!]
    \centering
    \includegraphics[width=\linewidth]{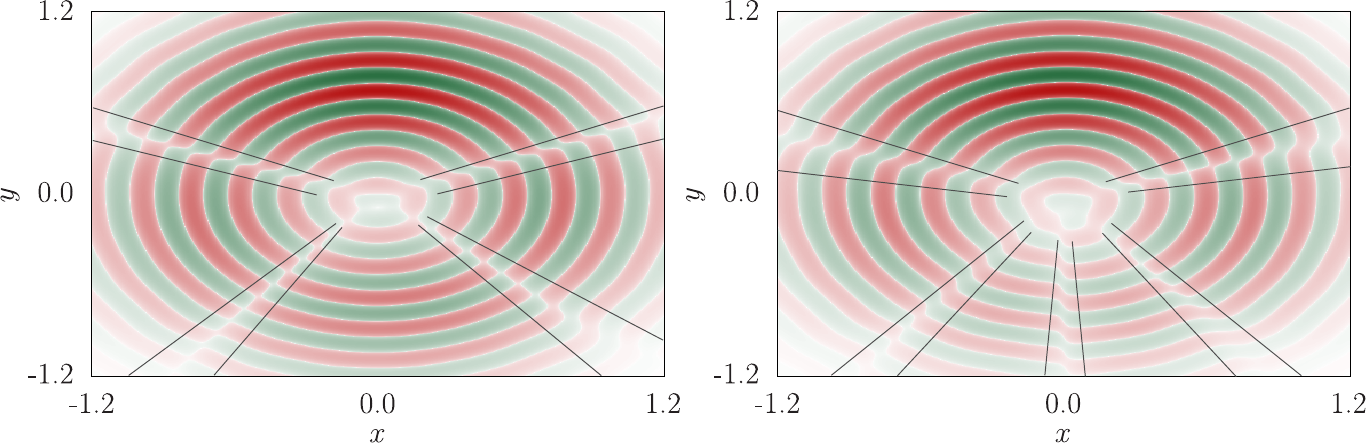}
    \put(-227,70){(\textbf{a})}
    \put(-18,70){(\textbf{b})}
    \caption{{\it Shot noise density-density correlations for two component fermions at intermediate interaction.} Cross-term correlator $\mathrm{Re}[G_{R,C}(\mathbf{x},\mathbf{x'};t)]$ for $N_{p} = 10$ fermions residing in a ring of $N_{s}=25$ sites with a repulsive interaction strength $U=50$. For a winding number \textbf{(a)} $\ell = 0$ there are four dislocations, whilst for $\ell = 1/2$ five dislocations appear, each marked by a pair of solid lines as a guide to the eye. In the plots obtained via DMRG, the Wannier functions are taken as two-dimensional time-evolved Gaussian functions with parameters radius $R=1.5$, Gaussian width $\sigma = 0.15$ and time of expansion $\omega_{0}t = 1.5$ fixing $\mathbf{x'} = (R,0)$. The color bar is taken to be non-linear by setting $\mathrm{sign}(Y)|Y)|^{\frac{1}{4}}$, with $Y$ denoting the quantity being plotted.}
    \label{fig:DMRG}
\end{figure}

{\textbf{Momentum distribution --}} Here, we show how dislocations can be explained through the features of momentum distribution: $n(k) = \sum_{j,l}e^{\imath k(j-l)}\langle c_{j}^{\dagger}c_{l}\rangle$, where $k$ denotes the lattice momenta (the  role of $n(k)$ in the interference protocols of fermionic gases was highlighted in~\cite{amico2005quantum,chetcuti2022interference}). 

For free and weakly interacting particles, $\ell$ can be represented by the occupation of the single-particle orbitals (such occupations provide indeed a partition of $\ell$). Clearly, while $n(k)$ is singly peaked for bosons, in the fermionic case $n(k)$ results to be flat around $k=0$. Essentially, dislocations emerge in the interference pattern  as an interference of the populated momenta of the system as $\frac{N_{p}}{2}-1$. For finite interaction, $n(k)$ broadens (with specific features depending on $\ell$) as shown in Sec.~\ref{sec:momdist} in Supplemental material, displaying a characteristic Fermi surface smearing (see f.i.~\cite{fradkin2013field}). We find that, whilst  single-particle levels are clearly not well defined anymore, dislocations  still arise from interference of sufficiently represented momenta in $n(k)$, with the aforementioned smearing possibly affecting  their visibility so (see Sec.~\ref{sec:momdist} in Supplemental material).

{\bf Conclusions and Outlook --} By combining rigorous analytical methods with numerical techniques, we proved that the zeros of the many-body wavefunction are marked by dislocations in the interference fringes arising from a protocol known as self-heterodyne phase reconstruction. This way, nodal surfaces show up as discontinuities of the phase, or phase slips~\cite{arutyunov2008superconductivity,d2017quantum,roscilde2016quantum}.
Irrespective of interaction, interference of bosonic systems in the ground-state give rise to continuous fringes, without dislocations. Symmetry-dictated nodal surfaces stem from the Pauli exclusion principle, whilst non-symmetry-dictated nodal surfaces~\cite{leggett1991theorem} come from interaction and depend on winding numbers characterizing the many-body wavefunction. Our theory includes the spin of the fermions: because of such degree of freedom, besides nodes, the many-body wavefunction can display characteristic cusps. The latter features can impart signature traits in the interference pattern as well. The  analysis of  the momentum distribution $n(k)$ indicates that the dislocations arise as {the} interference of the most populated quasi-momenta of the system, with such a property affecting the  visibility of the dislocations in the interference pattern.

Our study provides rigorous proofs on the structure of many-body wavefunction and demonstrates that its nodal surfaces are monitorable through quantum simulation. In cold atoms experiments with a small number of atoms, heterodyne interferograms can be analysed by image post-processing~\cite{delpace2022imprinting}. In this context, we mention the work carried out in ~\cite{brandstetter2025magnifying,lunt2024engineering,lunt2024realization} exploring degenerate atoms wavefunction with angular momentum in the deep mesoscopic limit (few particles).

Our work  provides a  valuable  window into the structure  of the many-body wavefunction. This way, we not only get insights into important chapters of quantum science  as Quantum Monte Carlo computational methods~\cite{ceperley1991fermion} and quantum phases of matter both at equilibrium~\cite{amico2008entanglement,eisert2010colloquium} and out of equilibrium~\cite{eisert2015quantum}, but we also demonstrate how important features of the wavefunction can be indeed explored through an interferometry-based quantum simulation. 

{\textbf{Acknowledgements --}} We would like to thank E.C. Domanti and A. Osterloh for fruitful discussions. WJC received funding from the European Union’s Horizon research and innovation programme under the Marie Sk\l odowska-Curie grant agreement \textit{SUN\textunderscore Atomtronics} (no. 101205763). AM acknowledges grants from projects Quantum-SOPHA ANR-21-CE47-0009 and Dyn1D  ANR-23-PETQ-0001. 

\bibliography{refs}

\begin{thebibliography}{55}
\bibitem [{Wan()}]{WannierGaussianNote}%
  \BibitemOpen
  \href@noop {} {}\bibinfo {note} {We utilize Wannier functions approximated as
  one-dimensional Gaussian functions with a width $\sigma = 0.4$.}\BibitemShut
  {Stop}%
\end{thebibliography}

\onecolumngrid
\setcounter{equation}{0}
\newpage

\section*{Supplementary Material}

\noindent In the following, we provide supporting details of the results found in the manuscript `\textit{Interferometric probe for the zeros of the many-body wavefunction}'.

\subsection{Details of the derivation of the free expansion of the many-body wavefunction}\label{sec:timeevol}

\noindent The time evolution of the many-body wavefunction of $N_{p}$ particles can be computed as
\begin{align}\label{eq:ogwave}
   \Psi (\mathbf{x}_{1},..,\mathbf{x}_{N_{p}};t) \!=\! \int \mathrm{d}\mathbf{x}_{1}'...\mathrm{d}\mathbf{x}_{N_{p}}'\mathcal{K}(\mathbf{x}_{1},..,\mathbf{x}_{N_{p}};t|\mathbf{x}_{1}',..,\mathbf{x}_{N_{p}}';0)\Psi (\mathbf{x}_{1}',..,\mathbf{x}_{N_{p}}';0),
\end{align}
via the Feynman kernel~\cite{feynman2005quantum} $\mathcal{K}(\mathbf{x}_{1},..,\mathbf{x}_{N_{p}};t|\mathbf{x}_{1}',..,\mathbf{x}_{N_{p}}';0) \!= \frac{1}{N_{p}!}\langle 0|\psi (\mathbf{x}_{N_{p}})...\psi (\mathbf{x}_{1})e^{-\imath\mathcal{H}_{0}t}\psi^{\dagger} (\mathbf{x}_{1}')...\psi^{\dagger} (\mathbf{x}_{N_{p}}') |0\rangle$, which is the many-body propagator~\cite{fetter2012quantum}. $\mathcal{H}_{0}=\sum_{\theta}^{N_{s}}a_{\theta}^{\dagger}a_{\theta+1}+a_{\theta+1}^{\dagger}a_{\theta}$ is the tight-binding Hamiltonian since we are dealing with a free expansion in time $t$ and $a^{\dagger}$ ($a$) denote the creation (annihilation) operators for bosons/fermions on site $\theta$. Expressing the field operators in terms of lattice operators through the Wannier basis $\psi^{\dagger}(\mathbf{x}) = \sum_{\theta=1}^{N_{s}}w^{*}(\mathbf{x}-\mathbf{R}_{\theta})a_{\theta}^{\dagger}$ where $w(\mathbf{x}-\mathbf{R}_{\theta})$ denotes a two-dimensional Wannier function localized around the lattice position $\mathbf{R}_{\theta}$, the propagator for $N_{p}$ particles reads
\begin{equation}
    \mathcal{K}(\mathbf{x}_{1},..,\mathbf{x}_{N_{p}};t|\mathbf{x_{1}'},..,\mathbf{x}_{N_{p}}';0) =  \sum\limits_{\theta_{1}...\theta_{N_{p}}}^{N_{s}}\sum\limits_{\vartheta_{1}...\vartheta_{N_{p}}}^{N_{s}}\Bigg[\prod\limits_{a=1}^{N_{p}}w(\mathbf{x}_{a}-\mathbf{R}_{\theta_{a}})\prod\limits_{b=1}^{N_{p}}w^{*}(\mathbf{x}_{b}'-\mathbf{R}_{\vartheta_{b}})\Bigg]\mathcal{K}_{N_{p}}^{\mathrm{lat}},
\end{equation}
where $\mathcal{K}_{N_{p}}^{\mathrm{lat}} = \frac{1}{N_{p}!}\langle 0|a_{\theta_{N_{p}}}..a_{\theta_{1}} e^{-\imath\mathcal{H}_{0}t}a_{\vartheta_{1}}^{\dagger}..a_{\vartheta_{N_{p}}}^{\dagger} |0\rangle$. Subsequently, for a non-interacting Hamiltonian such as $\mathcal{H}_{0}$, Wick's theorem can be applied that will either give a determinant $\mathcal{K}_{N_{p}}^{\mathrm{lat}} = \frac{1}{N_{p}!}\mathrm{det}[\mathcal{K}_{1}^{\mathrm{lat}}(\theta_{a};t|\vartheta_{b};0)]^{N_{p}}_{a,b=1}$ or permanent $\mathcal{K}_{N_{p}}^{\mathrm{lat}} = \frac{1}{N_{p}!}\mathrm{perm}[\mathcal{K}_{1}^{\mathrm{lat}}(\theta_{a};t|\vartheta_{b};0)]^{N_{p}}_{a,b=1}$ depending on whether the system is fermionic or bosonic respectively, where each entry in the matrix corresponds to the single-particle propagator $\mathcal{K}_{1}^{\mathrm{lat}}(\theta_{a};t|\vartheta_{b};0)$. \\

\noindent Defining the Fourier transform as $a_{\theta} = \frac{1}{\sqrt{N_{s}}}\sum_{k}e^{\imath k \theta} a_{k} $ with lattice momenta $k = \frac{2\pi}{N_{s}}n$ for $n=0,...,N_{s}-1$, we get that the Heisenberg equation in momentum space reads $\imath\hbar\frac{\mathrm{d}}{\mathrm{d}t}a_{k}(t) = [a_{k}(t),\mathcal{H}_{0}]$,where $\mathcal{H}_{0}=\sum_{k}\epsilon_{k}a_{k}^{\dagger}a_{k}$. Therefore, the single-particle lattice propagator, defined as $\mathcal{K}_{1}^{\mathrm{lat}}(\theta;t|\vartheta;0) = \langle 0|a_{\theta}(t)a_{\vartheta}^{\dagger}(0) |0\rangle$, has the following expression
\begin{equation}\label{eq:singleprop}
\mathcal{K}_{1}^{\mathrm{lat}}(j;t|j';0) = \frac{1}{N_{s}}\sum\limits_{k}^{N_{s}-1}e^{\imath k (j-j')}e^{-\imath \epsilon_{k}t/\hbar},
\end{equation}
which is the same expression as the single-particle Feynman kernel~\cite{feynman2005quantum}. Plugging everything back in Eq.~\eqref{eq:ogwave} we have that
\begin{equation}
    \Psi (\mathbf{x}_{1},..,\mathbf{x}_{N_{p}};t) \!=\! \frac{1}{N_{p}!}\int \mathrm{d}\mathbf{x}_{1}'...\mathrm{d}\mathbf{x}_{N_{p}}'\sum\limits_{\theta_{1}...\theta_{N_{p}}}^{N_{s}}\sum\limits_{\vartheta_{1}...\vartheta_{N_{p}}}^{N_{s}}\Bigg[\prod\limits_{a=1}^{N_{p}}w(\mathbf{x}_{a}-\mathbf{R}_{\theta_{a}})\prod\limits_{b=1}^{N_{p}}w^{*}(\mathbf{x}_{b}'-\mathbf{R}_{\vartheta_{b}})\Bigg] \mathcal{A}[\mathcal{K}_{1}^{\mathrm{lat}}(\theta_{c};t|\vartheta_{d};0)]\Psi (\mathbf{x}_{1}',..,\mathbf{x}_{N_{p}}';0),
\end{equation}
where $\mathcal{A}$ represents the determinant or the permanent. By noting that the wavefunction at $t=0$ is of the form $\Psi (\mathbf{x}_{1}',..,\mathbf{x}_{N_{p}}';0) = \sum_{j_{1}',...,j_{N_{p}}'}^{N_{s}}[\prod_{a=1}^{N_{p}}w(\mathbf{x}_{a}'-\mathbf{R}_{j_{a}})]\Psi (j_{1}',..,j_{N_{p}}';0)$, we can simplify the expression even further through the orthonormality of the Wannier functions such that 
\begin{equation}
    \Psi (\mathbf{x}_{1},..,\mathbf{x}_{N_{p}};t) \!=\! \frac{1}{N_{p}!}\sum\limits_{\theta_{1}...\theta_{N_{p}}}^{N_{s}}\sum\limits_{\vartheta_{1}...\vartheta_{N_{p}}}^{N_{s}}\Bigg[\prod\limits_{a=1}^{N_{p}}w(\mathbf{x}_{a}-\mathbf{R}_{\theta_{a}})\Bigg] \mathcal{G}[\mathcal{K}_{1}^{\mathrm{lat}}(\theta_{c};t|\vartheta_{d};0)]_{c,d=1}^{N_{p}}\Psi (\vartheta_{1},..,\vartheta_{N_{p}};0).
\end{equation}
For both fermions and bosons, the wavefunction’s symmetry is already encoded in its initial form, i.e. $\Psi (j_{1}',..,j_{N_{p}}';0)$. Thus, the many-body propagator simply evolves the underlying single-particle amplitudes and preserves that symmetry, rather than imposing it anew. In other words, the many-body propagator merely carries forward the single-particle amplitudes: the $N_{p}!$ identical contributions generated by the determinant/permanent cancel against the $\frac{1}{N_{p}!}$ prefactor to give 
\begin{equation}
    \Psi (\mathbf{x}_{1},..,\mathbf{x}_{N_{p}};t) \!=\! \sum\limits_{\theta_{1}...\theta_{N_{p}}}^{N_{s}}\Bigg[\prod\limits_{a=1}^{N_{p}}w(\mathbf{x}_{a}-\mathbf{R}_{\theta_{a}};t)\Bigg]\Psi (\theta_{1},..,\theta_{N_{p}};0),
\end{equation}
with the time dependence being absorbed in the Wannier functions $w(\mathbf{x}_{a}-\mathbf{R}_{\theta_{a}};t) \equiv \sum_{\vartheta_{a}}^{N_{s}}w(\mathbf{x}_{a}-\mathbf{R}_{\theta_{a}})\mathcal{K}_{1}^{\mathrm{lat}}(\theta_{a};t|\vartheta_{a};0)$. Approximating the Wannier functions as 2D Gaussians centered at $\mathbf{R}_{\theta_{a}}$ and performing the standard Gaussian integral, the time-dependence is encoded in the Wannier functions in the following way
\begin{equation}
    w(\mathbf{x}_{a}-\mathbf{R}_{\theta_{a}};t) = \frac{1}{\sigma\sqrt{\pi}}\frac{1}{1+\imath\omega_{0}t}\exp\bigg[-\frac{(\mathbf{x}_{a}-\mathbf{R}_{\theta_{a}})^{2}}{2\sigma^{2}(1+\imath\omega_{0}t)}\bigg],
\end{equation}
where frequency associated with the minimum of the lattice well in the harmonic approximation is $\omega_{0}=\hbar/(m\sigma^{2})$ with $\hbar$ being Planck's constant, $m$ the mass and $\sigma$ the width of the Gaussian.

\subsection{Explicit form of the bosonic/fermionic wavefunction}\label{sec:wavefunc}

\subsubsection{Non-interacting limit}\label{sec:zerointer}

\noindent For non-interacting particles, the wavefunction is built from single-particle orbitals $\varphi_{k}(\mathbf{x}) = \frac{1}{\sqrt{N_{s}}}\sum_{\theta}^{N_{s}}e^{\imath k \theta}w(\mathbf{x}-\mathbf{R}_{\theta})$ with $\theta$ being the lattice site on the one-dimensional ring and $k=\frac{2\pi n}{N_{s}}$ are the the lattice momenta with $n=0,...,N_{s}-1$. Depending on the nature of the particle statistics, the wavefunction can be obtained as a permanent or determinant of these single-particle orbitals $\Psi_{ring} (\{\mathbf{x}\})=
{\mathcal{A}} [ \varphi_{k_{q}}(\mathbf{x}_a)]_{q,a=1}^{N_{p}}$ for bosonic or fermionic systems respectively, with the (anti-)symmetrization denoted by $\mathcal{A}$ and $\{\mathbf{x}\} = \{ \mathbf{x}_{1},...,\mathbf{x}_{N_{p}}\}$ are the particle coordinates. \\

\noindent For bosons, we are interested in the situation where they all occupy in the same orbital as this gives the ground-state of the system. Thus, the wavefunction for these cases is simply
\begin{equation}
    \Psi^{B}_{ring}(\{\mathbf{x}\}) = \mathcal{N}_{B}\sum\limits_{\theta_{1},...,\theta_{N_{p}}}^{N_{s}}\bigg[\prod\limits_{a=1}^{N_{p}} w(\mathbf{x}_{a}-\mathbf{R}_{\theta_{a}})  e^{\imath k \theta_{a}}\bigg],
\end{equation}
where $\mathcal{N}_{B}$ is a normalization constant. On the other hand for fermions, the Pauli principle prohibits them from occupying the same orbital. Nonetheless, for the cases we are interested of a Slater determinant whose single-particle orbitals composing it have consecutive lattice momenta $k_{n}$, a nice clean expression can be obtained. \\

\noindent To start, let us consider $N_{p}$ spinless fermions and take the occupied momenta in $\varphi_{k}(\mathbf{x}_{j})$ to be consecutive integers $n = n_{0},n_{0}+1,...,,n_{0}+N_{p}-1$ with $n_{0}$ also being possibly negative as is the case for the Fermi sea distribution in the ground-state. Then, the orbitals are labeled by the index $q=0,...,N_{p}-1$ such that $k_{q} = \frac{2\pi}{N_{s}}(n_{0}+q)$. Consequently, the many-body wavefunction is written as
\begin{equation}
    \Psi_{ring}^{F}(\{\mathbf{x}\}) = \mathcal{N}_{F}\,\sum\limits_{\theta_{1},...,\theta_{N_{p}}}\mathrm{det}[e^{\frac{2\imath\pi}{N_{s}}(n_{0}+q)\theta_{a}}]\prod\limits_{a=1}^{N_{p}}w(\mathbf{x}_{a}-\mathbf{R}_{\theta_{a}}),
\end{equation}
where $\mathcal{N}_{F}$ is a normalization constant. Introducing $z_{a} := e^{\frac{2\imath\pi}{N_{s}}\theta_{a}}$ such that the matrix elements are $M_{qa} = z_{a}^{n_{0}+q}$. For each column $a$, we can take out the common factor $z_{a}^{n_{0}}$ to give
\begin{equation}
    \mathrm{det}[M] = \bigg(\prod\limits_{a=1}^{N_{p}}z_{a}^{n_{0}}\bigg)\mathrm{det}[z_{b}^{q}]_{b=1,...,{N_{p}}}^{q=0,...,{N_{p}-1}}.
\end{equation}
The remaining determinant with entries $z_{b}^{q}$ is a Vandermonde matrix~\cite{horn1985matrix} in the variables $\{z_{a}\}$ meaning that $\mathrm{det}[M] = \bigg(\prod_{a=1}^{N_{p}}z_{a}^{n_{0}}\bigg)\prod_{1\leq c<d}^{N_{p}}(z_{d}-z_{c})$. Putting everything together we have that 
\begin{equation}
    \Psi_{ring}(\{\mathbf{x}\}) = \mathcal{N}_{F}\sum\limits_{\theta_{1},...,\theta_{N_{p}}}^{N_{s}} \exp\bigg(\frac{2\imath\pi}{N_{s}}n_{0}\sum\limits_{l=1}^{N_{p}}\theta_{l}\bigg)\prod_{1\leq c<d\leq N_{p}}(z_{d}-z_{c})\prod\limits_{a=1}^{N_{p}}w(\mathbf{x}_{a}-\mathbf{R}_{\theta_{a}}).
\end{equation}
Lastly, we note that $e^{\gamma_{d}}-e^{\gamma_{c}} = e^{\imath\big[\frac{\gamma_{c}+\gamma_{d}}{2}\big]}2\imath\sin\left(\frac{\gamma_{d}-\gamma_{c}}{2}\right)$ with $\gamma_{w} = \frac{2\imath\pi}{N_{s}}\theta_{w}$. Collecting all the terms, we find that 
\begin{equation}\label{eq:noninter}
    \Psi_{ring}(\{\mathbf{x}\}) = \mathcal{D}_{F}\sum\limits_{\theta_{1},...,\theta_{N_{p}}}\left[ \exp\bigg(\frac{2\imath\pi}{N_{s}}k_{0}\sum\limits_{l=1}^{N_{p}}\theta_{l}\bigg)\prod_{1\leq c<d\leq N_{p}}\sin\left(\frac{\pi}{N_{s}}(\theta_{d}-\theta_{c})\right)\prod\limits_{a=1}^{N_{p}}w(\mathbf{x}_{a}-\mathbf{R}_{\theta_{a}})\right].
\end{equation}
with $\mathcal{D}_{F}$ being a normalization constant and $k_{0}=n_{0}+\frac{N_{p}-1}{2}$. Note that for the situation of consecutive momenta that we are considering $k_{0}$ can either be 0 or $\frac{1}{2}$ depending on whether the number of particles is odd or even. In the case of SU(2) fermions, the wavefunction is simply the tensor product of two chains of spinless fermions and can be built as the tensor product of Eq.~\eqref{eq:noninter}.

\subsubsection{Infinitely repulsive limit}\label{sec:infrep}

\noindent The presence of strong interactions in bosonic systems acts like an effective Pauli principle restraining them from residing in the same single-particle orbital. Therefore, the exact many-body wavefunction of $N_{p}$ bosons with hard-core interactions in the continuum is given by
\begin{equation}
    \Psi_\mathrm{TG}(\theta_{1},...,\theta_{N_{p}}) = \left[\prod\limits_{1\leq m<n\leq N_{p}} \mathrm{sign}(\theta_{n}-\theta_{m})\right]\mathrm{det}[\varphi_{k_{a}}(\theta_{j})]_{a,j=1}^{N_{p}}.
\end{equation}
Essentially, the Tonks-Girardeau wavefunction~\cite{girardeau1960relationship} is built from the Slater determinant of spinless fermions and the mapping function $\prod_{m<n}\mathrm{sign}(\theta_{n}-\theta_{m})$, which enforces the bosons' symmetry under two-particle exchange. For lattice systems, the wavefunction reads
\begin{equation}
    \Psi_\mathrm{TG}^{\mathrm{lat}}(\mathbf{x}_{1},...,\mathbf{x}_{N_{p}}) = \mathcal{N}_{\mathrm{TG}}\sum\limits_{\theta_{1},...,\theta_{N_{p}}}^{N_{s}}\left(\Bigg[\prod\limits_{1\leq m<n\leq N_{p}} \mathrm{sign}(\theta_{n}-\theta_{m})\Bigg]\mathrm{det}[e^{\imath k_{q}\theta_{b}}]_{b,q=1}^{N_{p}}\prod\limits_{a=1}^{N_{p}}w(\mathbf{x}_{a}-\mathbf{R}_{\theta_{a}})\right),
\end{equation}
where $\mathcal{N}_{\mathrm{TG}}$ is a normalization constant.  \\

\noindent Interacting two-component fermionic systems described by the Fermi-Hubbard model in the lattice and Gaudin-Yang Hamiltonian in the continuum are Bethe Ansatz integrable~\cite{lieb1968absence,gaudin1967un,yang1967some}. In a given sector $Q$, corresponding to the permutation of the relative ordering of the particle coordinates $\{\bar{\theta}\} = \theta_{1},...,\theta_{N_{p}}$, the wavefunction is of the form
\begin{equation}
    \Psi_{BA}^{F}(\{\bar{\theta}\},\{\bar{\alpha}\}|\bar{k},\bar{\lambda}) = \sum\limits_{P\in S_{N_{p}}}\mathrm{sign}(P)\mathrm{sign}(Q)\langle\bar{\alpha}Q|\bar{k}P,\bar{\lambda}\rangle\exp\left(\imath\sum\limits_{l=1}^{N_{p}} k_{Pl}\theta_{Ql}\right),
\end{equation}
with $\bar{k}$ being the charge quasimomenta, $\bar{\lambda}$ are the spin rapidities and $P$ denotes the permutation forming the symmetric group $S_{N_{p}}$. $\langle\bar{\alpha}Q|\bar{k}P,\bar{\lambda}\rangle$ is the spin wavefunction having a similar form to the eigenfunctions of the inhomogenous Heisenberg XXX model~\cite{deguchi2000thermodynamics}. In the limit of infinite repulsive interactions $U\rightarrow\infty$, the Bethe Ansatz equations for the charge and spin rapidities decouple~\cite{ogata1990bethe,essler2005one}. As a result, the corresponding wavefunction is of the form
\begin{equation}\label{eq:BAfermions}
    \Psi_{BA}^{\infty}(\{\bar{\theta}\},\{\bar{\alpha}\}|\bar{k},\bar{\lambda}) = \mathcal{B}_{F}\langle\bar{\alpha}Q|\bar{\lambda}\rangle\mathrm{det}\left[e^{\imath k_{m}\theta_{n}}\right]_{m,n=1}^{N_{p}},
\end{equation}
where $\mathrm{det}\left[e^{\imath k_{m}\theta_{n}}\right]$ is the Slater determinant of spinless fermions and $\mathcal{B}_{F}$ is a normalization constant. However, the momenta $k_{m}$ are not the lattice momenta. Instead, they are the charge rapidities obtained by solving the Bethe Ansatz equations that for $U\rightarrow\infty$ are expressed as $k_{l} = \frac{2\pi}{N_{s}}\left[I_{l}+\frac{1}{N_{p}}\sum_{m}^{M} J_{m}\right]$ for $l=1,...,N_{p}$ and $M$ being the number of down spins~\cite{osterloh2023exact}. The charge $I_{j}$ and spin $J_{m}$ quantum numbers characterise the spectrum of the system. In the presence of a flux $\phi$, the quantum number configurations can change to counterbalance and minimise the energy~\cite{yu1992persistent,chetcuti2022persistent}. For the cases we consider, the quantum number configurations adopted produce equally spaced $k_{l}$ allowing us to write the determinant in a similar form as Eq.~\eqref{eq:noninter} with the only difference being that the phase is given by $\exp\left(\imath \left[k_{0}+\frac{1}{N_{p}}\sum_{m}^{M}J_{m}\right]\sum_{l}^{N_{p}}\theta_{l}\right)$. On account of this decoupling, the spin wavefunction in the infinitely repulsive regime, is of the following form~\cite{essler2005one}
\begin{equation}\label{eq:spinwave}
    \langle\bar{\alpha}Q|\bar{\lambda}\rangle = \mathcal{N}\sum\limits_{P\in S_{M}}\left[ \prod\limits_{1\leq m<n\leq M} \frac{\Lambda_{P(m)}-\Lambda_{P(n)}-2\imath}{\Lambda_{P(m)}-\Lambda_{P(n)}}\right]\prod\limits_{l=1}^{M}\left(\frac{\Lambda_{P(l)}-\imath}{\Lambda_{P(l)}+\imath}\right)^{y_{l}},
\end{equation}
where $y_{l}$ is the coordinate of the fermions with spin down and the spin rapidities were re-scaled as $\lambda_{l} = \frac{U\Lambda_{l}}{4}$. The spin amplitudes no longer exhibit a dependence on the charge quasimomenta, which acted as the inhomogeneity in the eigenfunctions of XXX Hamiltonian. Indeed, to acquire the spin rapidities $\Lambda$ one needs to solve the Bethe equations of the Heisenberg XXX model
\begin{equation}
    \left(\frac{\Lambda_{l}-\imath}{\Lambda_{l}+\imath}\right)^{N_{p}} = \prod\limits_{m\neq l}^{M}\frac{\Lambda_{l}-\Lambda_{m}-2\imath}{\Lambda_{l}-\Lambda_{m}+2\imath} \hspace{5mm} l=1,...,M,
\end{equation}
which are parameterised by the same set of spin quantum numbers $\{J_{m}\}$ as the fermionic model to acquire its wavefunction. \\

\noindent For the present case of a fermionic ring pierced by an effective flux, $\{J_{m}\}$ will be modified to counteract the former. In doing so, the quasimomenta entering the Slater determinant are shifted by $\sum_{m}J_{m}$ and a specific Heisenberg state (corresponding to the spin amplitudes) is selected according to the configuration of the spin quantum numbers.

\subsection{From single-shot to expectation values}\label{sec:expect-single-shot}

\noindent In cold atoms experiments, self-heterodyne interferograms are obtained from a single run during a time-of-flight measurement as the quantum gases in the ring and center expand freely and interfere. Essentially, the interference picture emerges from the cross-term $\Psi_{1,ring}(\mathbf{x};t)\Psi_{1,center}^{*}(\mathbf{x};t)$ with $\Psi_{1,ring/center}$ denoting the one-particle wavefunction of the ring and center respectively.  For the ring, the one-dimensional wavefunction for one particle is defined as
\begin{equation}\label{eq:psione}
    \Psi_{ring}(\theta_{1}) = \sum\limits_{_{\theta_{1}=1}}^{N_{s}}\Psi_{ring}(\theta_{1},...,\theta_{N_{p}}),
\end{equation}
where $\Psi_{ring}(\theta_{1},...,\theta_{N_{p}})$ corresponds to the many-body wavefunction where $N_{p}-1$ particle coordinates ranging from $\theta_{2}$ to $\theta_{N_{p}}$ are fixed to a given position on the ring. A plot of $\Psi_{ring}(\theta_{1})$ in the ground-state as a function of $\theta_{1}$ displays $N_{p}-1$ zeros, either nodes or cusps, at the position of the fixed particle coordinates. Consequently, when looking at the interferogram generated from $\Psi^{*}_{1,ring}(\mathbf{x};t)\Psi_{1,center}(\mathbf{x};t)$ we observe dislocations (deformations) reflecting nodes (cusps) in the same positions as shown in Fig.~\ref{fig:singleshotnoninteracting} of the main text. Note that the nodes/cusps of the one-particle wavefunction in the continuum as per Eq.~\eqref{eq:psione} coincide with the ones in the lattice (acquired by multiplying the wavefunction by the one-dimensional Wannier function $w(x-\theta_{1})$ such that $\Psi_{ring}^{lat}(x) = \sum_{\theta_{1}}^{N_{s}}w(x-\theta_{1})\Psi_{ring}(\theta_{1},...,\theta_{N_{p}})$.\\

\noindent The same one-particle wavefunction can be identified in expectation values such as the density. By integrating out all the particle coordinates $\{\mathbf{x}_{j}\}$ in the plane but one such that
\begin{equation}
    n_{1}(\mathbf{x}) = \int \mathrm{d}\mathbf{x}_{2}...\mathrm{d}\mathbf{x}_{N_{p}}\Psi^{*}(\mathbf{x},\mathbf{x}_{2},...,\mathbf{x}_{N_{p}})\Psi(\mathbf{x},\mathbf{x}_{2},...,\mathbf{x}_{N_{p}}).
\end{equation}
Consider the non-interacting fermionic wavefunction in the lattice outlined in Eq.~\eqref{eq:noninter}, taking the Wannier functions to be two-dimensional Gaussians. Through the orthonormality of the Wannier functions $\int w^{*}(y-j)w(y-l)\mathrm{d}y = \delta_{jl}$, we find that
\begin{align}
    n_{1}(\mathbf{x}) = &\mathcal{D}_{F}\mathcal{D}^{*}_{F} \sum\limits_{\theta_{2},...,\theta_{N_{p}}}^{N_{s}} \left[\prod\limits_{2\leq p<q}^{N_{p}}\sin^{2}\left(\frac{\pi}{N_{s}}[\theta_{q}-\theta_{p}]\right)\right]\nonumber \\&\times\sum\limits_{\theta_{1},\vartheta_{1}}^{N_{s}}\exp\left(\frac{2\imath\pi}{N_{s}}k_{0}[\theta_{1}-\vartheta_{1}]\right)\prod\limits_{r = 1}^{N_{p}}\sin\left(\frac{\pi}{N_{s}}(\theta_{1}-\theta_{r})\right)\sin\left(\frac{\pi}{N_{s}}(\vartheta_{1}-\theta_{r})\right)  w^{*}(\mathbf{x}-\mathbf{R}_{\vartheta_{1}})w(\mathbf{x}-\mathbf{R}_{\theta_{1}}).
\end{align}
In the above equation, we separated the terms containing $\theta_{1}$ and $\vartheta_{1}$ from the rest. By doing so, the expression can be written to say that $n_{1}(\mathbf{x}) = \sum_{\{\tilde{\theta}\}}\Psi_{1,ring}^{*}(\mathbf{x})\Psi_{1,ring}(\mathbf{x})$ with $\Psi_{1,ring}(\mathbf{x}) = \sum_{\theta_{1}}^{N_{s}}w(\mathbf{x}-\mathbf{R}_{\theta_{1}})\Psi (\theta_{1},\{\tilde{\theta}\})$ and $\{\tilde{\theta}\}=\theta_{2},...,\theta_{N_{p}}$ allowing us to recognize that 
\begin{equation}\label{eq:oneparticle}
    \Psi (\theta_{1},\{\tilde{\theta}\}) = \mathcal{D}_{F}P(\theta_{2},...,\theta_{N_{p}})\exp\left(\frac{2\imath\pi}{N_{s}}k_{0}\theta_{1}\right)\prod\limits_{r=1}^{N_{s}}\sin\left(\frac{\pi}{N_{s}}(\theta_{1}-\theta_{r})\right),
\end{equation}
where $P^{*}(\theta_{2},...,\theta_{N_{p}}) P(\theta_{2},...,\theta_{N_{p}}) = \sin^{2}[\frac{\Delta k}{2}(\theta_{q}-\theta_{p})]$ is the probability of finding the other $N_{p}-1$ particles at a given point in space. From Eq.~\eqref{eq:one-particle-density}, the main quantity of interest is the zeros of the wavefunction that stem from the crossing of particle coordinates $\theta_{1}$ and $\theta_{r}$. So in the main text, we plot $\frac{\Psi(\theta_{1},\{\tilde{\theta}\})}{\mathcal{D}_{F}P(\theta_{2},...,\theta_{N_{p}})\exp\left(\frac{2\imath\pi}{N_{s}}k_{0}\theta_{1}\right)}$ in the top row of Fig~\ref{fig:singleshotnoninteracting}. \\

\noindent From a theoretical standpoint, the self-heterodyne interference patterns can be studied in the many-body formalism through the density-density correlator $G(\mathbf{x},\mathbf{x'},t) = \sum_{\alpha,\beta = \uparrow,\downarrow}\langle n_{\alpha}(\mathbf{x},t)n_{\beta}(\mathbf{x'},t)\rangle$ where $\alpha$ and $\beta$ are the colours/components of the system, $n_{\alpha}(\mathbf{x},t) = \Psi^{\dagger}_{\alpha}(\mathbf{x},t)\Psi_{\alpha}(\mathbf{x},t)$ and $\psi^{\dagger}_{\alpha}(\mathbf{x}) = (\psi^{\dagger}_{R,\alpha}(\mathbf{x})+\psi^{\dagger}_{C,\alpha}(\mathbf{x}))$ being the field creation operator for the full ring-center system denoted by subscripts $R$ and $C$ respectively~\cite{polo2025persistent}. Specifically, the interference emerges due to the ring-center cross terms and therefore we focus on $G_{R,C}(\mathbf{x},\mathbf{x'};t) = \rho_{1}^{R}(\mathbf{x},\mathbf{x'};t)\rho_{1}^{C}(\mathbf{x'},\mathbf{x};t)$ with $\rho_{1}^{R/C}(\mathbf{x},\mathbf{x'};t)$ being the one-body correlator~\cite{pecci2021probing,chetcuti2022interference}. Due to the ring and center systems being decoupled initially, their wavefunction can be viewed as a product state $|\Psi\rangle = |\Psi_{R}\rangle\otimes|\Psi_{C}\rangle$ allowing us to separate the ring and center terms in the $G_{R,C}(\mathbf{x},\mathbf{x'};t)$. In first quantization, the one-body correlator reads:
\begin{equation}\label{eq:corrdef}
    \rho_{1}^{R/C}(\mathbf{x},\mathbf{x'}) \!=\! \int \mathrm{d}\mathbf{x}_{2}...\mathrm{d}\mathbf{x}_{N_{p}}\Psi^{*}(\mathbf{x},\mathbf{x}_{2},...,\mathbf{x}_{N_{p}})\Psi(\mathbf{x'},\mathbf{x}_{2},...,\mathbf{x}_{N_{p}}).
\end{equation}
Let us consider non-interacting fermions and focus on the ring correlator. Plugging the expression for the wavefunction in Eq.~\eqref{eq:corrdef} we find that
\begin{align}\label{eq:correxpnoninter}
    \rho_{1}^{R}(\mathbf{x},\mathbf{x'}) = &\mathcal{D}_{F}\mathcal{D}_{F}^{*}\sum_{\theta_{2},...,\theta_{N_{p}}}\left[\prod\limits_{2\leq p<q}^{N_{p}}\sin^{2}\bigg[\frac{\Delta k}{2}(\theta_{q}-\theta_{p})\right] \nonumber \\
    \times &\sum_{\theta_{1},\vartheta_{1}}\exp\left(\frac{2\imath\pi}{N_{s}} k_{0}(\theta_{1}-\vartheta_{1})\right)\prod\limits_{q=2}^{N_{p}}\sin\left[\frac{\Delta k}{2}(\theta_{1}-\theta_{q})\right]\sin\bigg[\frac{\Delta k}{2}(\vartheta_{1}-\theta_{q})\bigg]\times w^{*}(\mathbf{x'}-\mathbf{R}_{\vartheta_{1}})w(\mathbf{x}-\mathbf{R}_{\theta_{1}})\Bigg].
\end{align}
This can then be arranged and recast to give $\rho_{1}^{R}(\mathbf{x},\mathbf{x'}) = \sum_{\{\tilde{\theta}\}}^{N_{s}}\Psi _{1,ring}^{*}(\mathbf{x'})\Psi_{1,ring}(\mathbf{x})$ with $\Psi_{1,ring}(\mathbf{x})$ as defined in Eq.~\eqref{eq:oneparticle}. The dislocations present in interferograms correspond to zeros in the one-body correlator. Through Eq.~\eqref{eq:correxpnoninter} we understand that these zeros come from $\Psi_{1,ring}(\mathbf{x})$ as it changes sign whenever the first particles encounters the position of the other ones. Note that to get the final observable of $\rho_{1}^{R}(\mathbf{x},\mathbf{x'})$, one needs to average over the positions of all the particles with their probability $P(\theta_{2},...,\theta_{N_{p}})$, whilst $\Psi^{*}_{1,ring}(\mathbf{x'})$ is not generating further zeros in $\mathbf{x'}$. \\

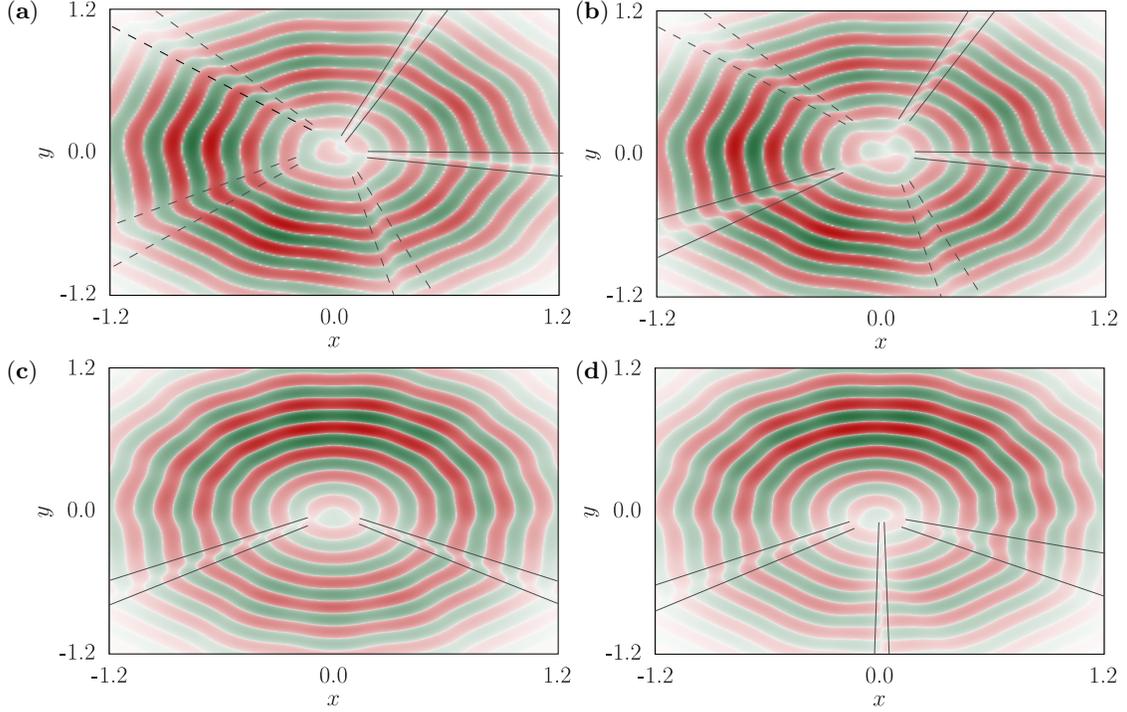
\begin{figure}[h!]
    \centering
    \includegraphics[width=0.8\linewidth]{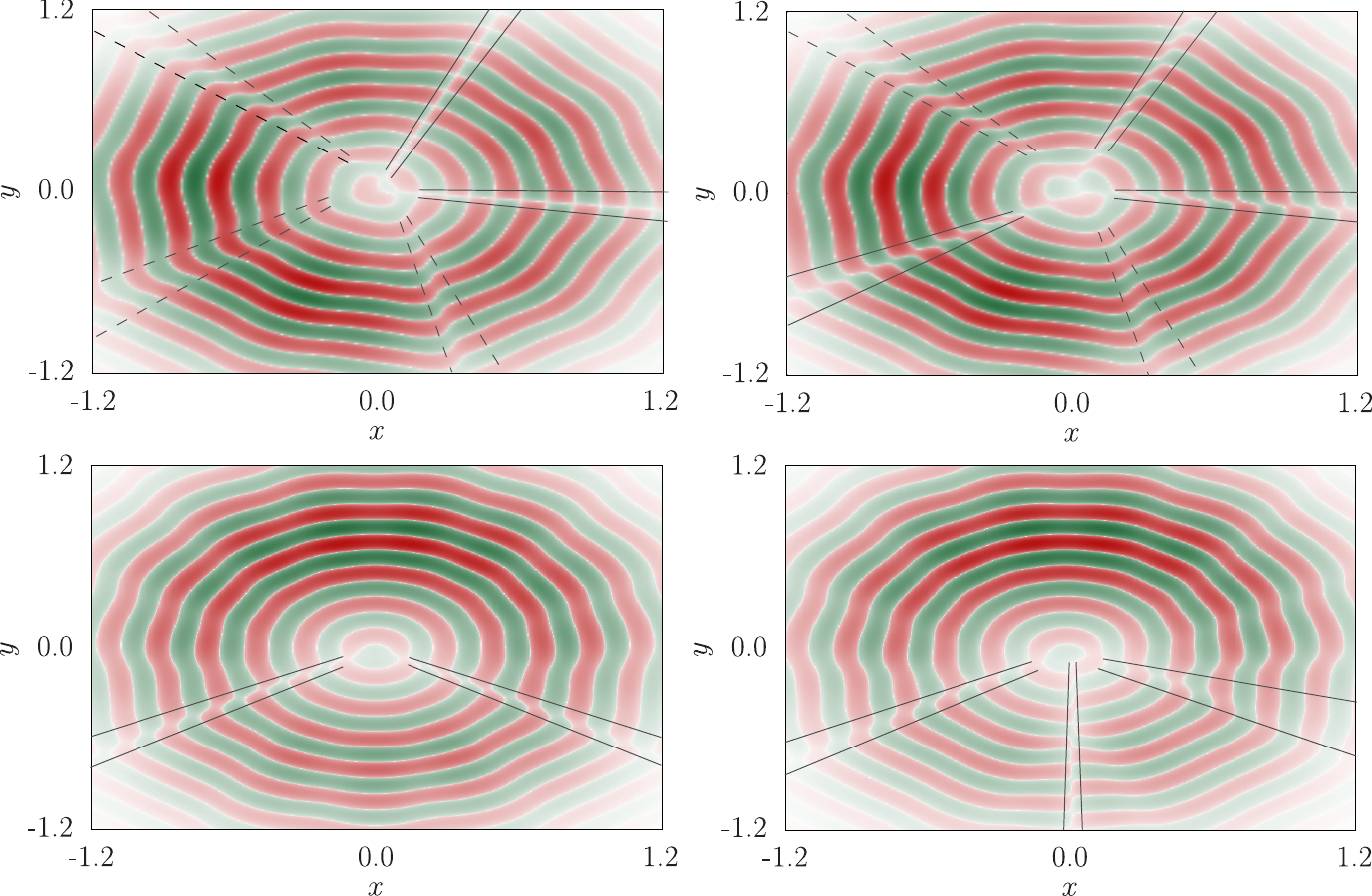}
    \put(-420,260){(\textbf{a})}
    \put(-205,260){(\textbf{b})}
    \put(-420,124){(\textbf{c})}
    \put(-205,124){(\textbf{d})}
    \caption{{\it{Comparison of the interference patterns acquired from the single-shot scheme} $\mathrm{Re}[\Psi^{*}_{1,center}(\mathbf{x};t)\Psi_{1,ring}(\mathbf{x};t)]$ (top row) and the cross-term correlator $\mathrm{Re}[G_{R,C}(\mathbf{x},\mathbf{x'};t)]$ (bottom row).} For the plots,  $N_{p} = 6$ two-component fermions residing in a ring of $N_{s}=15$ sites are considered. In both protocols, we observe that the number of dislocations coincides with two being present when $\mathbf{(a)},\textbf{(c)}$ $\ell=0$ case and three for $\mathbf{(b)},\mathbf{(d)}$ $\ell=\frac{1}{2}$. For the single-shot protocol, we utilize $\Psi_{1,ring}(\mathbf{x};t)/(\mathcal{D}_{F}P(\theta_{2},...,\theta_{6}))$ defined in Eq.~\eqref{eq:oneparticle} with the particle positions fixed as $\{\theta_{2},\theta_{3},\theta_{4},\theta_{5},\theta_{6}\} = 1, 4, 7, 10, 13$. Dislocations are denoted by solid lines whilst cusps are marked by dashed lines. There are two (three) dislocations and three (two) cusps in the single-shot scheme for $\ell=0$ ($\ell = \frac{1}{2}$) respectively. In the plots, the Wannier functions are taken as two-dimensional time-evolved Gaussian functions with parameters radius $R=1.5$, Gaussian width $\sigma = 0.15$ and time of expansion $\omega_{0}t = 1.5$. To generate the plots in the bottom row for the full cross-term correlator, we utilized exact diagonalization fixing $\mathbf{x'} = (R,0)$. The color bar is taken to be non-linear by setting $\mathrm{sign}(Y)|Y)|^{\frac{1}{4}}$, with $Y$ denoting the quantity being plotted.}
    \label{fig:comparison}
\end{figure}

\noindent The same procedure of extracting the one-particle wavefunction from observables can also be applied for the bosonic wavefunction and in the infinitely interacting regime. However, for two-component fermions in the limit $U\rightarrow\infty$ the one-particle wavefunction as expressed in Eq.~\eqref{eq:BAfermions} introduces some caveats. To start we re-visit the single-shot scheme for fermions in this limit. The spin wavefunction as defined in Eq.~\eqref{eq:spinwave} is dependent on the coordinate of the down spin fermions $y_{l}$ and in turn on the spin configuration of the particles. To acquire the single-shot interferogram, we fix the initial spin configuration to $\big|\!\uparrow_{1}...\uparrow_{\frac{N_{p}}{2}}\downarrow_{\frac{N_{p}}{2}+1}...\downarrow_{N_{p}} \big\rangle$, associating the position of the free particle $\theta_{1}$ with $\uparrow_{1}$ and fixing the rest. As $\uparrow_{1}$ makes its way across the ring, there will be different particle coordinate sectors $Q$. Specifically, for $N_{p}$ particles with an equal number per colour there are $\frac{N_{p}!}{\frac{N_{p}}{2}!\frac{N_{p}}{2}!}$. For the single-shot scheme, since we are fixing the particle positions we do not go through all the sectors. As an example, for $N_{p}=6$ particles starting from the initial spin configuration $|\!\uparrow\uparrow\uparrow\downarrow\downarrow\downarrow\rangle$
and ending at $|\!\uparrow\uparrow\downarrow\downarrow\downarrow\uparrow\rangle$, we pass through 4 sectors: $\{|\!\uparrow\uparrow\uparrow\downarrow\downarrow\downarrow\rangle,|\!\uparrow\uparrow\downarrow\uparrow\downarrow\downarrow\rangle,|\!\uparrow\uparrow\downarrow\downarrow\uparrow\downarrow\rangle,|\!\uparrow\uparrow\downarrow\downarrow\downarrow\uparrow\rangle\}$. However, when the interferograms are obtained through the correlator, all coordinate sectors contribute. Consequently, the dislocations in the resulting interference pattern need not appear as sharply as in the single-shot scheme. In the single-shot protocol, nodes give rise to dislocations and cusps produce mild distortions; in contrast, with the correlator the nodes lead to dislocations or weaker deformations, while the cusps are largely washed out -- Fig.~\ref{fig:comparison}. \\

\noindent Lastly, we note that taking advantage of the spin-charge decoupling in the wavefunction at $U\rightarrow\infty$, the fermionic one-body correlator can be cast in the following form~\cite{ogata1990bethe,osterloh2023exact}
\begin{equation}
    \langle c_{l}^{\dagger}c_{j} \rangle = \sum\limits_{x}\Psi_{charge}^{*}(l,x)\Psi_{charge}(j,x) S(j'\rightarrow l')
\end{equation}
where $x = \{x_{2},...,x_{N_{p}}\}$ with $x\subset \{1,...,N_{s}\}\setminus\{j,l\}$, $\Psi_{charge}$ denotes the Slater determinant before and after the particle hops from positions $j$ to $l$, and $S(j'\rightarrow l')$ is the corresponding product of the spin wavefunctions~\cite{ogata1990bethe,osterloh2023exact}
\[
S(j'\rightarrow l') = \bigg\{
\begin{aligned}
   \langle P_{l',l-1}...P_{j'+1,j'}\rangle_{H_{\rm{XXX}}}, \hspace{5mm}j'<l'\\
   \langle P_{l',l+1}...P_{j'-1,j'}\rangle_{H_{\rm{XXX}}},\hspace{5mm}j'>l'
\end{aligned}
\]
where $j'$ and $l'$ is the position of the $j'$th and $l'$th spins in the Heisenberg chain, $P_{q,r}$ is the permutation operator exchanging spins $q$ and $r$ evaluated on the Heisenberg XXX eigenstate associated to the given choice of Bethe Ansatz spin quantum numbers $\{J_{m}\}$.

\subsection{Relation between momentum distribution and self-heterodyne protocol}\label{sec:momdist}

\noindent The ring-center cross terms giving rise to the interference can be re-cast in the following form~\cite{pecci2022single,chetcuti2022interference}
\begin{equation}
    G_{R,C}(\mathbf{x},\mathbf{x'};t) = \frac{1}{N_{s}}\sum\limits_{\alpha = \uparrow,\downarrow}\sum\limits_{j,l}^{N_{s}}I(\mathbf{x};t)I^{*}(\mathbf{x'};t)\langle a_{j\alpha}^{\dagger}a_{l\alpha}\rangle,
\end{equation}
where we used the relation $\langle a_{0}^{\dagger}a_{0}\rangle = 1$ for the correlator of the central site, $N_{s}$ is the number of sites, $\alpha$ denotes the component number and defined
\begin{equation}
    I(\mathbf{x}) = \Bigg|\frac{1}{\sigma\sqrt{\pi}}\frac{1-\imath\tau}{b^{2}(\tau)}\Bigg|^{2}\exp\left[-\frac{\mathbf{x}^{2}}{2\sigma^{2}b^{2}(\tau)}\right]\exp\left[-\frac{\imath\tau}{2\sigma^{2}b^{2}(\tau)}\mathbf{x}^{2}\right]\sum\limits_{j=1}^{N_{s}}\exp\left[-\frac{(\mathbf{x}-\mathbf{x}_{j})^{2}}{2\sigma^{2}b^{2}(\tau)}\right]\exp\left[\frac{\imath\tau}{2\sigma^{2}b^{2}(\tau)}(\mathbf{x}-\mathbf{x}_{j})^{2}\right],
\end{equation}
setting $\tau = \omega_{0}t$ and $b(\tau) = \sqrt{1+\omega_{0}^{2}t^{2}}$. \\

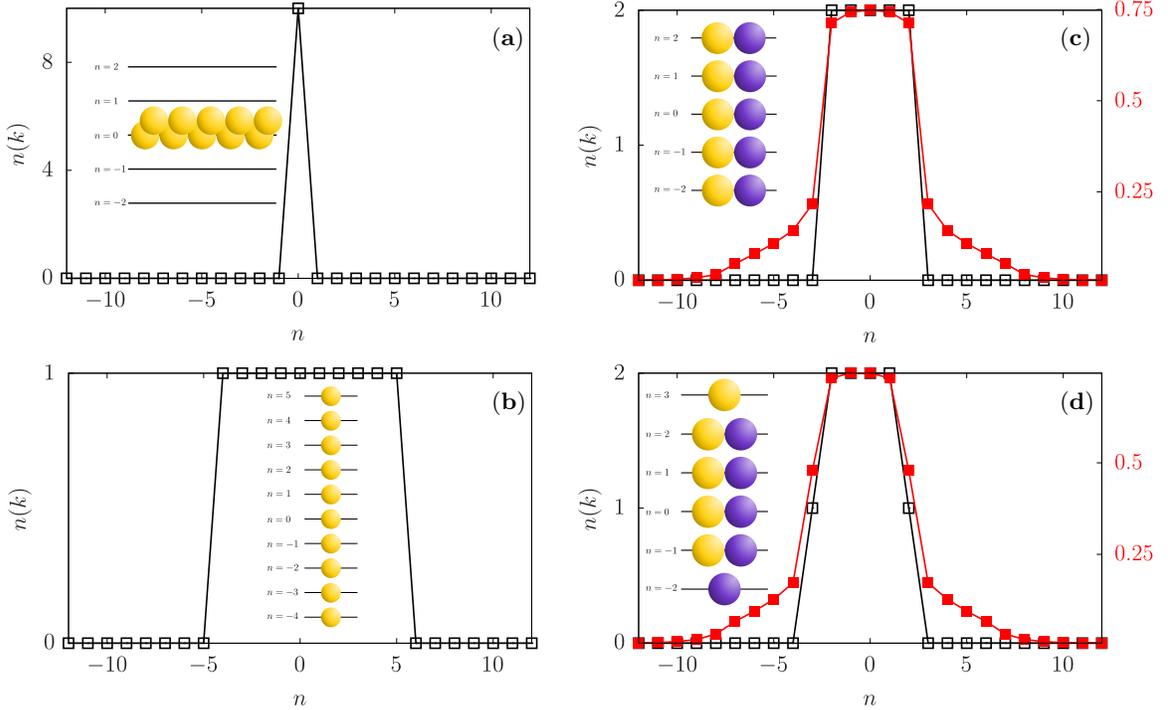
\begin{figure}[htbp!]
    \centering
\includegraphics[width=0.85\linewidth]{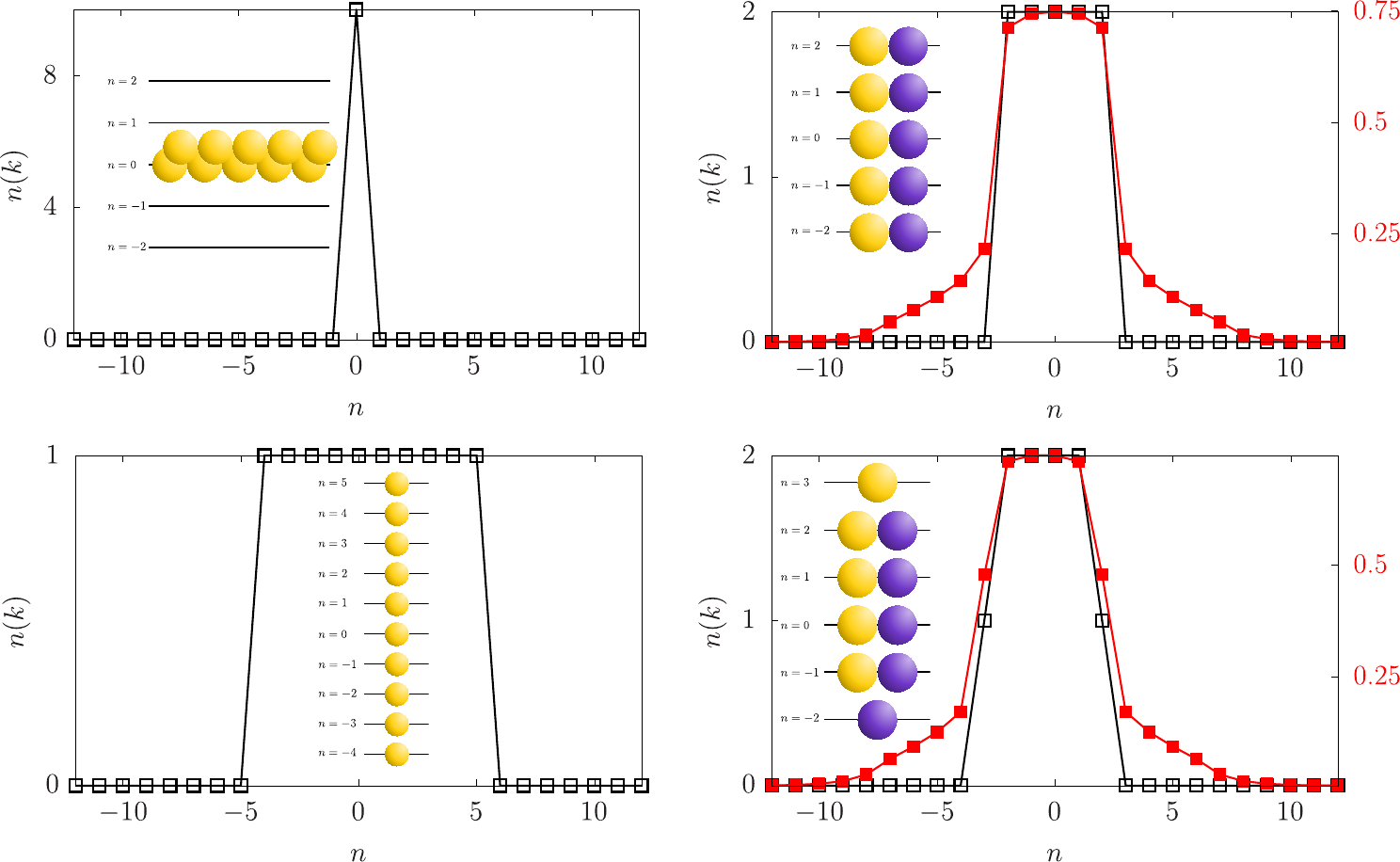}%
    \put(-250,250){(\textbf{a})}
    \put(-35,250){(\textbf{c})}
    \put(-250,112){(\textbf{b})}
    \put(-35,112){(\textbf{d})}
    \caption{\textit{Momentum distribution}. $n(k)$ for $N_{p}=10$ $\mathbf{(a)}$ free bosons, $\mathbf{(b)}$ spinless fermions and $\mathbf{(c)},\mathbf{(d)}$ two-component fermions. Plots \textbf{(c)} and \textbf{(d)} depict both the non-interacting (black) and interacting (red) cases for winding number $\ell = 0$ and $\ell=1/2$. Schematic depicts the momentum occupation of the corresponding winding number $\ell$ at zero interactions. Note that the configuration of momenta for the $\ell = 1/2$ case is degenerate with the one where the yellow and purple colours are swapped. Smearing and widening of the momentum distribution reducing visibility of zeros. Plots were obtained with DMRG for a ring composed of $N_{s}=25$ sites and interaction strength $U=50$.
    }
\label{fig:MomentumDistribution}
\end{figure}

\noindent The one-body correlator can be expressed in terms of the momentum distribution via 
\begin{equation}\label{eq:momdist}
   \langle c_{j\alpha}^{\dagger}c_{l\alpha}\rangle  = \frac{1}{N_{s}}\sum\limits_{k}^{N_{s}-1}e^{-\imath k (j-l)}n_{\alpha}(k),
\end{equation}
where $k$ denotes $n(k)$ is the momentum distribution of a given lattice momenta $k$. For $U=0$, we have that $n_{\alpha}(k) =1$, such that 
\begin{equation}
    G_{R,C}(\mathbf{x},\mathbf{x'};t) = \frac{1}{N_{s}}\sum\limits_{\alpha=\uparrow,\downarrow}\sum\limits_{j,l}^{N_{s}}I(\mathbf{x};t)I^{*}(\mathbf{x'};t)\sum\limits_{\{n\}}e^{-\frac{2\imath\pi}{N_{s}}n(j-l)},
\end{equation}
with $k=\frac{2\pi n}{N_{s}}$ and $n$ being the quantum number labeling the energy levels occupation. In the ground-state and zero flux, all the bosons occupy the same $k=0$ whilst fermions are compactly distributed around $k=0$. When a flux threads the system, the momenta occupation shifts to oppose it. At any interaction $U$,
\begin{equation}
    G_{R,C}(\mathbf{x},\mathbf{x'};t) = \frac{1}{N_{s}}\sum\limits_{\alpha=\uparrow,\downarrow}\sum\limits_{j,l}^{N_{s}}I(\mathbf{x};t)I^{*}(\mathbf{x'};t)\sum\limits_{\{n\}}e^{-\imath\frac{2\pi}{N_{s}}n(j-l)}n(k).
\end{equation}
The momentum distribution $n(k)$ being obtained from numerics as the Fourier transform of the one-body correlator as in Eq.~\eqref{eq:momdist}. In Fig.~\ref{fig:MomentumDistribution}, we show the momentum distribution for $N_{p}=10$ particles to highlight its broadening with increasing interaction. Whilst going to larger interactions introduces smearing, it is important to note that the lattice momenta present in the non-interacting picture are still the main contributors to the interference and the ones that dictate both the dislocation number and their visibility. 

\subsection{Parameters for DMRG}\label{sec:numerics}

\noindent For the DMRG simulations in Fig.~\ref{fig:DMRG} of the main text, we employed 300 sweeps with a maximum bond dimension of 4000 in the presence of finite flux, while 100 sweeps were sufficient for simulations without an imposed flux. The fermionic system is modeled as two interacting species of spinless fermions with periodic boundary conditions. Each physical lattice site is represented by two spinless sites in an interleaved one-dimensional chain, with one species assigned to even sites and the other to odd sites, such that a single physical site corresponds to an odd–even pair. In this representation, on-site interspecies interactions are mapped to nearest-neighbor interactions along the effective chain, while conserving the total particle number. We find that this site-doubling formulation leads to significantly faster convergence to the ground-state in DMRG compared to the standard spinful fermion implementation in ITensor.jl. Although the interleaved formulation doubles the number of lattice sites, it reduces the local Hilbert space and maps on-site interspecies interactions to nearest-neighbor couplings. This converts local (on-site) entanglement into spatial entanglement, which better matches the variational structure of DMRG and results in faster convergence. We remark that periodic boundary conditions, together with the degeneracies of the strongly interacting regime, limit the applicability of DMRG; in these cases, we resort to analytical methods.

\end{document}